\documentclass[journal=jacsat,manuscript=article]{achemso}
\SectionNumbersOn

\usepackage[version=3]{mhchem} 
\usepackage{booktabs}
\usepackage{tabularx}
\usepackage{makecell}
\usepackage{adjustbox}
\usepackage{rotating}
\usepackage{color,soul}
\usepackage{subfigure}
\usepackage{caption}
\usepackage{array}
\usepackage{amssymb}
\usepackage{xr}
\usepackage{xr-hyper}
\usepackage{hyperref}
\usepackage{subcaption}
\usepackage{braket}
\usepackage{multirow}
\usepackage{graphicx}
\usepackage{achemso}
\usepackage{natbib}
\usepackage{pdflscape}
\setkeys{acs}{etalmode = truncate}
\setkeys{acs}{maxauthors = 10}

\author{Younghwan Kim}
\affiliation[University of Chicago]{Department of Chemistry, University of Chicago, Chicago, IL 60637, United States}
\altaffiliation{Contributed equally to this work}
\author{Ming-Hsiu Hsieh}
\affiliation[University of Chicago]{Department of Chemistry, University of Chicago, Chicago, IL 60637, United States}
\altaffiliation{Contributed equally to this work}
\author{Ankit Mathanker}
\affiliation[University of Chicago]{Department of Chemistry, University of Chicago, Chicago, IL 60637, United States}
\author{Arpan Choudhury}
\affiliation[University of Chicago]{Department of Chemistry, University of Chicago, Chicago, IL 60637, United States}
\author{Joshua Zhou}
\affiliation[University of Chicago]{Pritzker School of Molecular Engineering, University of Chicago, Chicago, IL 60637, United States}
\author{Qiaohong Wang}
\affiliation[University of Chicago]{Pritzker School of Molecular Engineering, University of Chicago, Chicago, IL 60637, United States}
\author{Matthew R. Hermes}
\affiliation[University of Chicago]{Department of Chemistry, University of Chicago, Chicago, IL 60637, United States}
\author{Laura Gagliardi}
\affiliation[University of Chicago]{Department of Chemistry, University of Chicago, Chicago, IL 60637, United States}
\alsoaffiliation[University of Chicago]{Pritzker School of Molecular Engineering, University of Chicago, Chicago, IL 60637, United States}

\email{lgagliardi@uchicago.edu}

\title[]
{Autonomous Multireference Quantum Chemistry with Artificial Intelligence Agents}

\begin{document}

\begin{abstract}
Accurately describing chemical reactions that involve strongly correlated electrons remains a major challenge in computational chemistry. Multireference electronic structure methods can address this challenge, but their application requires expert decisions that are difficult to automate, particularly when the electronic structure changes as chemical bonds form and break. Here we introduce CASPER, an autonomously agentic workflow that performs multireference calculations along chemical reaction pathways. 
CASPER combines chemical knowledge, electronic structure calculations, and iterative assessment to identify chemically meaningful active spaces and maintain their consistency as bonds form and break. We demonstrate its capabilities on five transition-metal reactions, for which CASPER generates consistent multireference potential energy profiles with minimal human intervention. Importantly, in two reactions, CASPER identifies critical electronic-structure problems that an independent Claude session fails to recognize without the structured workflow. These include an electronic-state crossing and the exclusion of an orbital involved in bond breaking. Our results show the importance of combining language models with chemical knowledge and independent assessment of computational results. Beyond automating calculations, such agents have the potential to help researchers better understand the underlying physics and chemistry and open new opportunities for investigating complex chemical reactivity and generating high-quality quantum chemical data at scale.
\end{abstract}

\section{Introduction}
Understanding how chemical bonds form and break requires an accurate description of the electronic structure. This is particularly challenging in transition-metal complexes, where several electronic configurations can contribute to the wavefunction and their relative importance can change significantly along a reaction pathway.\cite{wardzala2026oxo,ghosh2006transition}

Multireference electronic structure methods provide a powerful framework for describing these strongly correlated systems,  \citep{wardzala2026multireference,vitillo2022multireference,choudhury2024understanding,mandal2025modeling} but their application remains far from routine. One of the main challenges is selecting an appropriate active space, namely the subset of electrons and orbitals treated explicitly in the multiconfigurational wavefunction. This selection requires considerable expertise and often repeated manual intervention.\cite{veryazov2011how} The problem becomes particularly challenging for chemical reactions, where the active space must provide a balanced description of the evolving electronic structure as bonds form and break.

Considerable progress has been made towards automating active space selection. Methods based on orbital entanglement, such as autoCAS,\cite{stein2016automated,stein2019autocas} approximate correlation measures,\cite{king2021ranked,king2022large,wardzala2025automated} machine learning,\cite{golub2021machine,jeong2020automation,osaro2026rlease} and atomic orbital projections\cite{kolodzeiski2023automated,tarocco2025aegiss}, including atomic valence active space (AVAS)\cite{sayfutyarova2017automated} and $\pi$-orbital system (PiOS)\citep{sayfutyarova2019constructing}, have been developed. Other methods, including the weighted active space protocol (WASP),\cite{seal2025weighted} orbital mapping strategies,\cite{bensberg2023corresponding} and adaptive orbital selection methods\cite{li2026self} address the consistency of active spaces across molecular geometries. More recently, large language model (LLM) agents have been introduced to assist with electronic structure calculations and active space selection.\cite{lee2026autonomous,perezsanchez2026elagente,li2026automatedmanybody} 

However, identifying an active space at a single geometry is only part of the problem. During a chemical reaction, orbitals change their character, electronic states may become nearly degenerate, and chemically important orbitals can be replaced during complete active space self-consistent field (CASSCF) optimization. Consequently, calculations may converge and produce smooth energy profiles without providing a chemically appropriate description of the reaction. Recognizing these problems and deciding how to correct them requires more than automating calculations. It requires chemical knowledge, analysis of the electronic structure, and the ability to revise computational decisions.

Here we introduce CASPER (Complete Active Space Planning, Execution, and Reasoning), an artificial intelligence agentic framework for autonomous multireference calculations along chemical reaction pathways. CASPER combines chemical knowledge with electronic structure calculations in a structured workflow that selects active spaces, evaluates their chemical character and consistency, identifies potential problems, and modifies the calculations when necessary. We demonstrate CASPER on five transition-metal reactions involving different bond-forming and bond-breaking processes. In all five cases, CASPER identifies consistent active spaces and generates smooth potential energy profiles with minimal human intervention.

An important result of this work is the comparison between CASPER workflow and an independent Claude session. In the two reactions examined, controlled Claude session was able to perform the calculations and obtain apparently consistent active spaces. However, it failed to identify an electronic-state crossing in one reaction and excluded an orbital involved in bond breaking in another. CASPER recognized these problems and modified the calculations accordingly. Although this comparison is limited to two reactions, it shows that convergence and orbital consistency alone are not sufficient to establish the chemical validity of multireference calculations. It also suggests that a structured workflow with specialized chemical assessments can improve the quality of the results obtained with an LLM.

Beyond automating electronic structure calculations, we envision that agents such as CASPER can help scientists gain a better understanding of the physics and chemistry of the systems they investigate. By analyzing the electronic structure, identifying possible problems, and suggesting how to address them, these agents can complement human expertise and reduce the time spent on technical aspects of the calculations. This would allow researchers to focus more on interpreting the results and addressing important scientific questions. CASPER represents an initial step towards making multireference calculations more accessible and useful for studying complex chemical reactivity.

\section{An Agentic Framework for Autonomous Multireference Calculations}
We developed CASPER, an agentic framework that integrates chemical reasoning, electronic structure calculations and iterative assessment to automate multireference calculations along chemical reaction pathways. The central challenge is not simply to identify an appropriate active space at a given molecular geometry, but to ensure that the selected orbitals retain their chemical relevance and provide a consistent description of the evolving electronic structure throughout the reaction. CASPER addresses this challenge by combining automated computational procedures with specialized committee members that evaluate different aspects of the calculations and recommend corrective actions when necessary.

\begin{figure}[t!]
  \includegraphics[width=\textwidth]{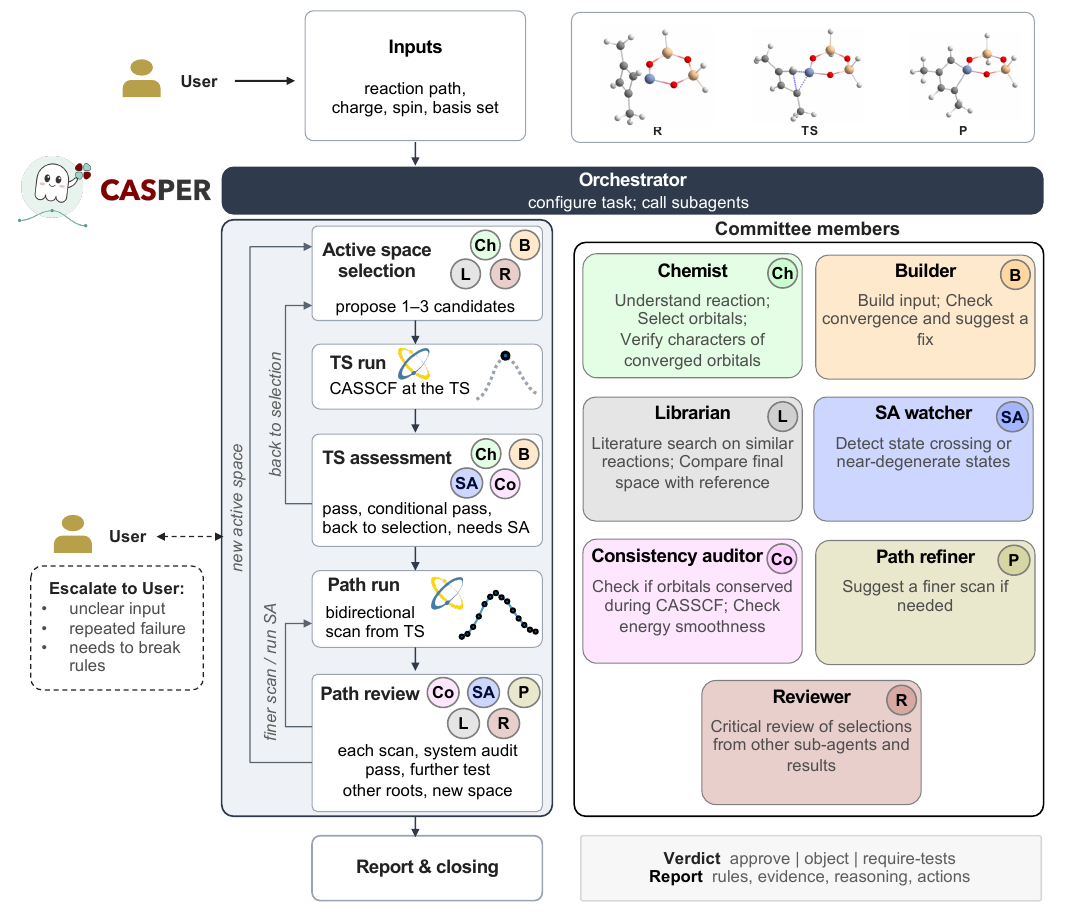}
  \caption{Overview of the agentic workflow for active space selection for a reaction path. Workflow requires reaction path, system spin, charge, and basis set as inputs. The main agent session (\textbf{Orchestrator}) carries one reaction path through five stages: active space selection, transition-state (TS) run, TS assessment, path run, path review, followed by report generation. During each stage, the workflow can access multiple committee members, one at a time, namely: \textbf{Chemist}, \textbf{Builder}, \textbf{Librarian}, \textbf{Reviewer}, \textbf{Consistency Auditor}, \textbf{Path Refiner}, and \textbf{SA Watcher}. The \textbf{Orchestrator} briefs the relevant members of the committee and receives a structured verdict. The \textbf{Librarian} and \textbf{Reviewer} run as fresh sub-agents that do not see other agents' reasoning. The user is consulted only when the committee cannot proceed.}
  \label{fig_workflow}
\end{figure}

The framework is organized into five stages: (1) active space selection, (2) transition-state (TS) calculation (TS Run), (3) TS assessment, (4) reaction-path calculation (Path Run), and (5) path review (Figure \ref{fig_workflow}). The workflow begins at a reference geometry, typically the TS, where chemically motivated candidate active spaces are constructed and state-specific complete active space self-consistent field (SS-CASSCF) calculations are performed and evaluated. Once a suitable active space has been identified, calculations are propagated along the reaction coordinate in both directions. Throughout this process, CASPER monitors the chemical character of the active orbitals, the consistency of the optimized active spaces and changes in electronic-state character. When inconsistencies are detected, the framework can try different initial guesses, revise the orbital selection or refine the reaction path before continuing. The workflow ends only after the resulting calculations satisfy the prescribed consistency criteria and undergo a final independent evaluation.

Scientific decisions are coordinated by an \textbf{Orchestrator} that consults seven specialized committee members (Figure ~\ref{fig_workflow}). The \textbf{Chemist} interprets the reaction chemistry and identifies orbitals relevant to bond formation, bond cleavage, and metal--ligand interactions. The \textbf{Builder} constructs computational inputs, executes calculations, monitors convergence and proposes remedies when calculations fail. The \textbf{Consistency Auditor} evaluates whether the optimized active orbitals retain their intended chemical character and remain consistent across molecular geometries. The \textbf{State-average (SA) Watcher} monitors the electronic states and identifies situations in which state-averaged calculations are required. The \textbf{Path Refiner} determines whether inconsistencies arise from insufficient geometric resolution and recommends additional intermediate structures when necessary. The \textbf{Librarian} examines relevant literature to identify comparable electronic structures and active spaces, while the \textbf{Reviewer} independently evaluates the scientific validity of the proposed selections and computational results.

Each committee member operates according to a defined set of instructions and assessment criteria, returning a structured recommendation to the \textbf{Orchestrator}. The members do not communicate directly with one another, allowing their assessments to be considered separately. When recommendations conflict, the \textbf{Orchestrator} evaluates the available computational evidence and, when necessary, initiates additional calculations to resolve the disagreement. While most members are roles that the \textbf{Orchestrator} applies in turn within its own context, the \textbf{Librarian} and \textbf{Reviewer} are always invoked with fresh, isolated contexts as subagents during the final assessment to independently scrutinize the results. This organization separates the generation of computational decisions from their critical evaluation, reducing reliance on any single member's assessment.

An important feature of CASPER is that convergence of an electronic structure calculation does not, by itself, constitute success. A converged CASSCF calculation may contain orbitals that have changed their chemical character or have been replaced during orbital optimization, potentially compromising the consistency of the reaction profile. CASPER therefore evaluates the optimized active space using quantitative orbital-overlap measures, natural orbital occupations, and chemically informed analysis of orbital character. These assessments determine whether a calculation can be accepted or whether further refinement is required. The framework also monitors changes in electronic-state character and can introduce state averaging when appropriate. By combining these diagnostics with iterative decision-making, CASPER can identify and correct problems that would otherwise require expert intervention.

The workflow is governed by explicit computational constraints and escalation rules. Human input is requested only when the initial specification is ambiguous, repeated failures cannot be resolved within the established rules, or a proposed action would exceed the permitted computational scope. All decisions, supporting evidence, and corrective actions are recorded, providing a traceable account of how the final active space was obtained. This combination of specialized scientific reasoning, independent assessment, and computational feedback provides the foundation for autonomous multireference calculations of chemical reactivity.

\subsection{Input and Preparation}
CASPER requires a series of molecular geometries describing a reaction pathway, together with the basis set, molecular charge, spin multiplicity, and the geometry at which the calculations should begin. In the present work, we use the transition-state (TS) geometry obtained from the reference density functional theory (DFT) calculations as the starting point. Although the user can select any geometry, we refer to the starting geometry as the TS throughout the manuscript.

The molecular geometries are first aligned along the reaction pathway. A restricted Hartree–Fock or restricted open-shell Hartree–Fock calculation is then performed at the TS, depending on the spin state. The occupied orbitals are localized using intrinsic bond orbitals (IBOs) \citep{knizia2013intrinsic} and the valence virtual orbitals are localized using localized intrinsic valence virtual orbitals (LIVVOs).\citep{schmidt2015valence, derricotte2017localized} Each orbital is characterized by its intrinsic atomic orbital (IAO) populations, resolved by atom and atomic shell, its bonding character ($\sigma$, $\pi$ or $\delta$), and its bonding or antibonding partner. Orbitals are also grouped into conjugated $\pi$ and sets of metal d orbitals.

The metal oxidation state and d-electron count at the TS are assigned using localized-orbital bonding analysis (LOBA),\citep{thom2009LOBA} and IAO-based bond orders are calculated. Changes in bonding along the reaction pathway are identified from the molecular geometries and classified as bond formation, bond cleavage, changes in bond order, transient contacts near the TS, or changes in metal–ligand distances without bond formation or cleavage. This information provides the chemical basis for the initial active space selection.

\subsection{Active Space Selection}

The initial active space is generated automatically at the TS (or any starting frame specified by the user) using the IBO/LIVVO procedure. The minimal active space includes singly occupied orbitals, five metal d orbitals, and the orbitals involved in bond formation and cleavage along the reaction path. Additional orbitals are ranked according to their chemical relevance, giving priority to covalent metal--ligand orbitals and their correlating partners, followed by orbitals associated with transient or secondary metal--ligand interactions and conjugated $\pi$ orbitals involving the reacting atoms. These additional orbitals are considered for expanding the active space but are not included by default. The \textbf{Chemist} examines the molecular structures and reaction path considering the metal oxidation state, the characters of orbitals, and bond events, and proposes an active space based on a predefined chemical checklist. The \textbf{Builder} compares this proposal with the automatically generated selection and constructs one to three candidate active spaces, generally ranging from a minimal selection to larger spaces containing additional chemically relevant orbitals. When the IBO/LIVVO procedure does not yield an unambiguous orbital selection, the \textbf{Builder} can instead use the AVAS approach. 
To limit computational cost, the \textbf{Builder} is restricted to active spaces containing no more than 18 electrons in 18 orbitals, (18e, 18o), \emph{i.e.}~$2.4\times 10^9$ determinants. Active spaces corresponding to a single determinant are also excluded. At the same time, the \textbf{Librarian} searches the literature and reports chemically related active spaces. The \textbf{Reviewer} evaluates the proposed active spaces and determines which candidates should proceed to CASSCF calculations.

\subsection{Transition State (TS) Run and Assessment}
SS-CASSCF calculations are performed at TS for the selected candidate active spaces using the spin multiplicity and basis set specified by the user.
After CASSCF convergence, multiconfiguration pair-density functional theory (MC-PDFT)\cite{li2014multiconfiguration} calculations with the tPBE and tPBE0 functionals and a three-root state-averaged complete active space configuration interaction (SA-CASCI) calculation are carried out. 
Each calculation is analyzed using an automated assessment script that examines convergence, spin contamination, natural orbital occupation numbers (NOONs), leading electronic configurations, and the chemical character of the natural orbitals (NOs), determined from meta-L\"owdin atomic and shell populations. \citep{lowdinNonOrthogonalityProblemConnected1950} The script also monitors whether the orbitals selected in the initial active space remain within the optimized active space.
The \textbf{Consistency Auditor} first compares the converged NOs with the initial guess using the orbital similarity metric, defined as the mean singular value 
    \begin{equation}
s(\mathbf{C}_1,\mathbf{C}_2) = \frac{1}{n}\sum_{k=1}^{n}\sigma_k\!\left(\mathbf{C}_1^{\mathsf{T}}\,\mathbf{S}\,\mathbf{C}_2\right),
    \label{eq_SV}
    \end{equation}
where $\mathbf{C}_1$ and $\mathbf{C}_2$ are the coefficient matrices for the two sets of $n$ active orbitals, and $\mathbf{S}$ is the atomic overlap matrix. $\sigma_k$ are the singular values of the overlap matrix. 
A similarity $s$ below 0.80 indicates a substantial change in the active orbital subspace and triggers further examination for orbital substitutions. 
The \textbf{Consistency Auditor} also identifies orbitals with core or Rydberg character that may have entered the active space during optimization.

The \textbf{Chemist} examines these results to determine whether the optimized active space retains the orbitals needed to describe the reaction. Particular attention is given to orbitals involved in bond formation and cleavage and to metal d orbitals. For each NO with a NOON above 1.995 or below 0.005, the \textbf{Builder} checks whether the corresponding correlating orbital is included in the active space. If a partner orbital is missing, the \textbf{Builder} either adds it or removes
the nearly doubly occupied or empty orbital and construct a revised active space from the converged NOs. The \textbf{SA Watcher} examines the spin purity and leading configurations of each state and identifies possible state crossings or strong state mixing. When the first excited state lies within 5.0 kcal/mol of the ground state and satisfies the prescribed weight criterion of 0.1, the \textbf{SA Watcher} initiates state-averaged CASSCF or linearized pair-density functional theory (L-PDFT) calculations.\cite{hennefarth2023linearized} Otherwise, the calculation remains state-specific.
Based on these assessments, the TS calculation receives one of four outcomes: acceptance, conditional acceptance with additional checks required along the reaction pathway, rejection followed by a new active space selection, or a recommendation to perform state-averaged calculations.

\subsection{Path Run and Review}
Once a suitable active space has been identified at the TS, CASSCF calculations are propagated along the reaction pathway in both directions, towards the reactant and product. At each geometry, CASSCF is followed by MC-PDFT, or L-PDFT for state-averaged calculations, and a three-root SA-CASCI calculation. The initial orbitals for each geometry are obtained by projecting the converged natural orbitals from the preceding geometry and orthonormalizing them in the new atomic orbital basis.
The quality of the projected orbitals is evaluated using the similarity metric $s$ defined in Eq.~\ref{eq_SV}. For comparisons between different geometries, the cross-geometry overlap matrix is used. During CASSCF optimization, an automated script monitors the similarity between the initial and current active space every five macroiterations, beginning at iteration 10. The calculation is stopped if the similarity falls below 0.90 at two consecutive checks. The \textbf{Builder} also monitors the calculations for convergence problems and terminates jobs that have stalled or failed.

Once the Path Run have reached both ends of the reaction pathway, or have stopped because of convergence or orbital-consistency problems, the results are examined by the \textbf{Consistency Auditor}, \textbf{SA Watcher}, and \textbf{Path Refiner}. The \textbf{Consistency Auditor} compares the optimized active spaces at neighboring geometries using the orbital similarity metric and the chemical character of the natural orbitals. A reaction pathway passes the consistency assessment when the similarity between all neighboring active spaces exceeds 0.97. When this criterion is not satisfied, the orbitals responsible for the change are identified and included in the assessment report.
The \textbf{SA Watcher} monitors changes in NOONs and leading configuration interaction coefficients to determine whether the same electronic state is followed along the reaction pathway. When a state crossing or significant state mixing is detected, additional SA-CASSCF tests are performed and state-averaged calculations are introduced when necessary. The \textbf{Path Refiner} determines whether an inconsistency may arise from an excessively large geometrical change between neighboring structures. Additional geometries are recommended when the change in a bond length exceeds 0.3 $\text{\AA}$ and the projected initial orbitals have low similarity with those at the preceding geometry. These intermediate structures are generated using image-dependent pair potential (IDPP) interpolation.\cite{smidstrup2014improved} The framework therefore does not require a uniformly fine reaction pathway, but introduces additional geometries only when needed to improve active space consistency.

If any of the three committee members rejects the results, the \textbf{Orchestrator} initiates another reaction-path calculation following their recommendations. Once the initial assessment is complete, the \textbf{Consistency Auditor}, \textbf{Librarian}, and \textbf{Reviewer} perform a final audit using fresh, isolated contexts to reduce the influence of earlier decisions. The \textbf{Librarian} compares the selected active space with those reported for related systems, the \textbf{Reviewer} evaluates its chemical validity, and the \textbf{Consistency Auditor} double-checks the consistency of the active spaces and computational settings across the reaction pathway. The final assessment can lead to acceptance, additional tests, a change in the electronic states considered, or a new active space selection. In the latter case, the workflow restarts with a new run identifier.

\subsection{Report and Closing}
Throughout the workflow, the \textbf{Orchestrator} records the chemical characterization of the selected orbitals, the decisions made by each committee member, the criteria used to justify these decisions, and the results of successive assessments. A separate record is maintained for each active space trial, allowing the selection process and any corrective actions to be traced.
Once the final active space has been accepted, CASPER generates an interactive report containing the selected active space, molecular geometries and natural orbitals at each point along the reaction pathway, absolute and relative energies obtained with the different electronic structure methods, and the complete decision history. Molecular structures and orbitals can be inspected as three-dimensional representations.

\section{Demonstration of CASPER and Results}
We tested CASPER on five organometallic reactions selected from the Gold-DIGR dataset\citep{li2026mining}. The reactions, shown in Table \ref{Table_reaction_structures}, involve different bond-breaking and bond-forming processes and provide a test of the ability of CASPER to identify consistent active spaces along reaction pathways.
Reaction A involves a doublet \ce{O-O} bond formation between a bound benzoate and the oxo in an iron-oxo bis(picolinate) benzoate complex in the doublet state.\citep{li2025peroxide} Reaction B involves a hydrogen transfer from 1-hexene to form chromacycloheptane with a bis(pyridyl)amine ligand.\citep{liu2019what} 
Reaction C describes the ring expansion of a coordinated $\eta^4$-dimethylcyclobutadiene to chromacyclopentadiene.\citep{liu2013reactivity} Reaction D involves the loss of propyne from a chromacycloheptatriene complex.\citep{liu2013reactivity} Reaction E is similar to Reaction B but involves a thioether-amine ligand and hydrogen transfer from chromacycloheptane to 1-hexene.\citep{yang2011spin} Reactions B--E are in the triplet state.

The reactant and product structures for each reaction were reoptimized at the PBE0-D4/cc-pVDZ level using the def2/J auxiliary basis set under the Resolution of the Identity for the Coulomb term combined with the Chain-Of-Spheres Exchange method (RIJCOSX), and \texttt{TightSCF} convergence criteria. Climbing-image nudged elastic band (NEB) calculations were performed at the same level using 12 images, followed by TS optimization starting from the climbing image. When the optimized TS structure deviated from the climbing image, the NEB trajectory was refined with the optimized TS held fixed. We limited our current tests to reactions that involve a single transition metal and are single-barrier without spin crossing. Additional single-point calculations at the $\omega$B97M-V/cc-pVDZ level with the same density fitting and convergence criteria were performed to obtain reference energies. The DFT and NEB calculations were carried out with ORCA 6.0.1. \citep{neese2012orca, neese2025software}

\begin{table}[htbp]
\caption{Five test organometallic reactions and their spin multiplicity.}
\label{Table_reaction_structures}
\centering
\renewcommand{\arraystretch}{1.5}
\begin{tabular}{c c c}\hline
ID & Formula & Spin multiplicity \\ \hline
A & \includegraphics[width=10cm,height=1.7cm,keepaspectratio]{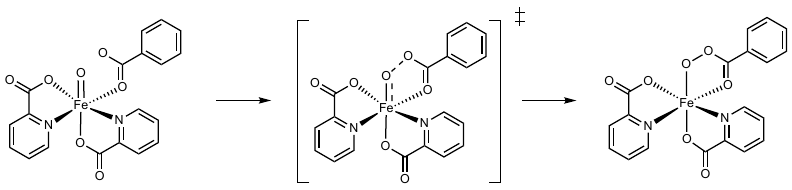} & 2 \\ 
B & \includegraphics[width=10cm,height=1.7cm,keepaspectratio]{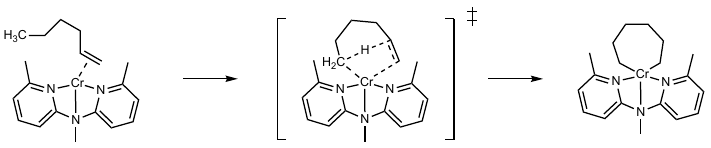} & 3 \\ 
C & \includegraphics[width=10cm,height=1.7cm,keepaspectratio]{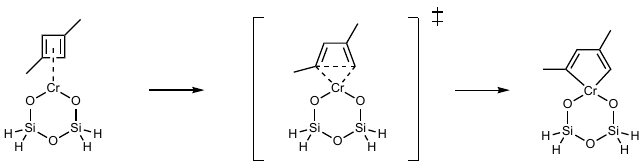} & 3 \\ 
D & \includegraphics[width=10cm,height=1.7cm,keepaspectratio]{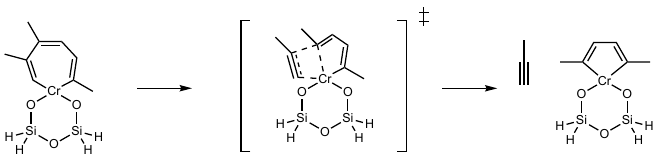} & 3 \\ 
E & \includegraphics[width=10cm,height=1.7cm,keepaspectratio]{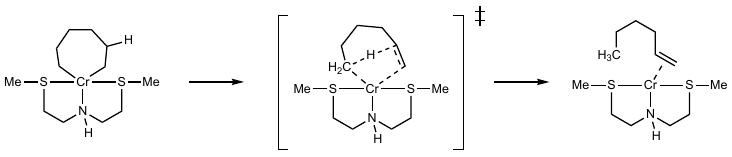} & 3 \\ \hline
\end{tabular}
\end{table}

CASPER was run with Claude Opus 5.5 with Medium effort. SS- and SA-CASSCF, MC-PDFT, L-PDFT, and strongly contracted second-order $n$-electron valence state perturbation theory (NEVPT2), are carried out with PySCF 2.14\cite{sun2020recent,sun2026python} and PySCF-forge 1.1.1\cite{pyscf-forge}. GPU-accelerated code is from gpu4mrh\cite{mrh}. All calculations are carried out with the cc-pVDZ basis set, along with the def2-universal-jkfit auxiliary basis for density fitting. The NEVPT2 calculation is not included in the CASPER workflow but is shown in our results to provide an energy reference.

Table \ref{table_SVD} summarizes the active spaces selected by CASPER, the number of trials required, and whether SS- or SA-CASSCF was used. Active space consistency was evaluated using the orbital similarity metric defined in Eq.~\ref{eq_SV}, calculated between the converged active orbitals of neighboring geometries. CASPER identified consistent active spaces for all five reactions, with mean similarities ranging from 0.9946 to 0.9991. The number of trials varied from four for Reaction B to sixteen for Reaction A. The final active spaces were also examined by the authors, and the corresponding natural orbitals are presented in Supplementary Figures S3--S8. In four of the five reactions, SS-CASSCF was sufficient. Reaction C required SA-CASSCF with two roots because of a crossing between electronic states of the same spin multiplicity along the reaction pathway.

We discuss Reactions C and E in more detail because they illustrate two different challenges in active space selection and how CASPER addresses them.

\begin{table} 
\caption{Selection results of each reaction.} 
\label{table_SVD} 
\centering 
    \begin{tabularx}{0.75\textwidth}{c c c c c}
    \hline Reaction & \# of trials & SS or SA & $(N_\mathrm{e}, N_\mathrm{o})$ &  $s$ (mean $\pm$ std) \\
    \hline 
    A & 16 & SS & (11, 11) & $0.9991 \pm 0.0028$  \\
    B & 4 & SS & (10, 10) & $0.9970 \pm 0.0028$ \\
    C & 6 & SA-2 & (10, 11) & $0.9957 \pm 0.0020$ \\
    D & 10 & SS & (14, 16) & $0.9946 \pm 0.0085$ \\
    E & 5 & SS & (10, 10) & $ 0.9977 \pm 0.0017$ \\
    \hline 
    \end{tabularx}
\end{table}

\subsection{Reaction C: Detecting and Addressing State Crossing}

\begin{figure}[t!]
  \includegraphics[width=0.9\textwidth]{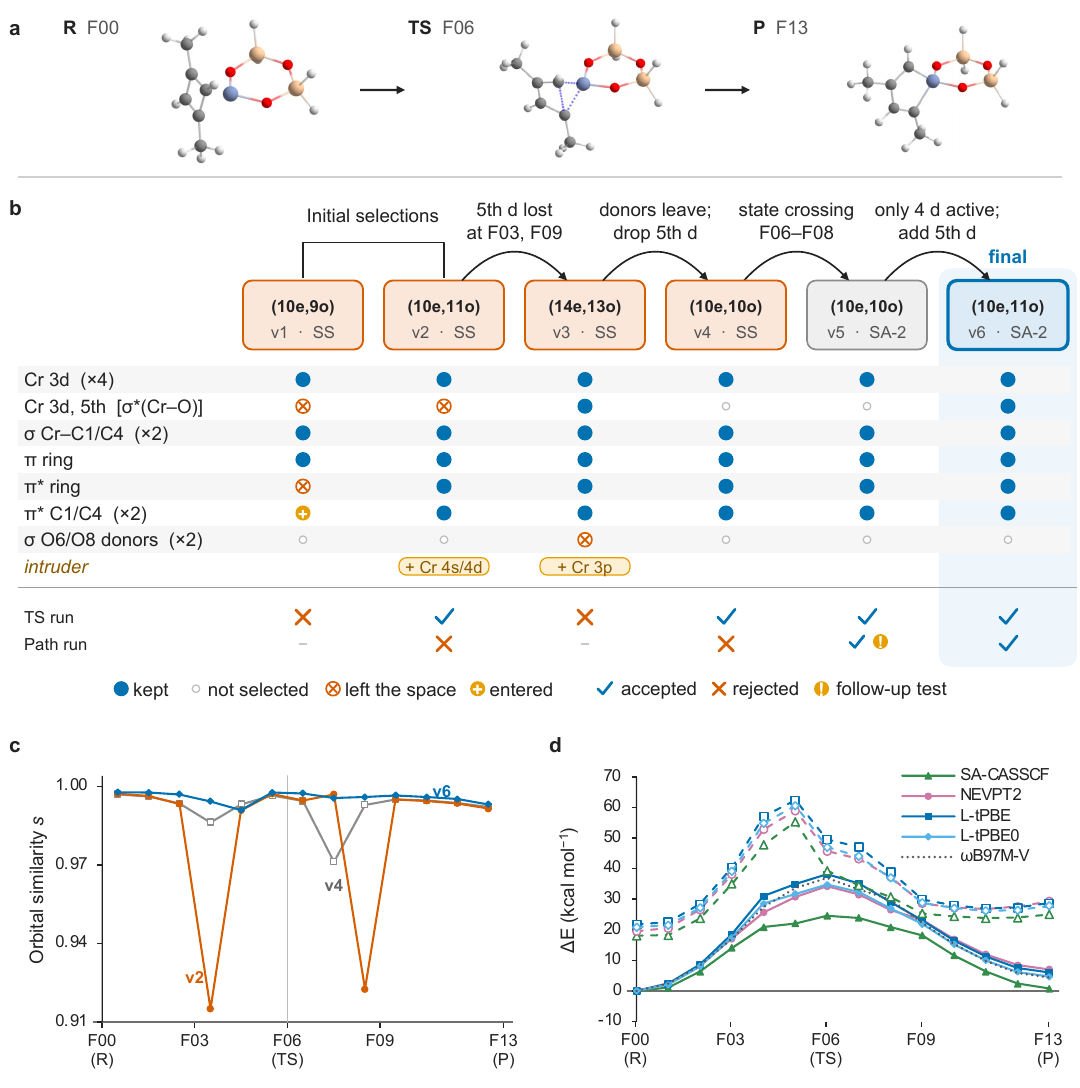}
  \caption{Step-by-step active space selection process along the path of Reaction C using CASPER. (a) Geometries for reactant, TS, and product, (b) The six trial active spaces in the order the workflow selects them. Each orbital row shows whether they are kept, left, or entered the active space during either TS or Path Run for each trial. The bottom two rows show the verdicts of the TS Run and the Path Run. (c) Orbital similarity  between neighboring frames $s$ for the trials v2, v4 and v6. (d) Energy profile of the adopted (10e, 11o) relative to reactant (F00) at SA-CASSCF, NEVPT2, L-PDFT (tPBE, tPBE0), along with a respective $\omega$B97M-V result. SS, state-specific; SA-$2$, state-averaged over the $2$ lowest triplet roots (solid and dashed, respectively) with equal weights.}
  \label{fig_rxnC}
\end{figure}

Reaction C provides an example in which more than one active spaces are accepted at the TS, but state crossing occurs, and thus the state-average treatment is required to capture consistent evolution of electronic states (Figure \ref{fig_rxnC}). The initial selection was a (10e,9o) active space containing the five Cr 3d orbitals, the \ce{Cr-C} $\sigma$ and $\sigma^*$ orbitals involved in bond cleavage, and a $\pi/\pi^*$ pair associated with the cyclobutadiene ring. The \textbf{Chemist} proposed including two additional $\pi^*$ orbitals, leading to a (10e,11o) active space. Both selections were tested at the TS (frame 06, F06).
The (10e,9o) active space failed because a Cr 3d orbital and a $\pi^*$ orbital were replaced during CASSCF optimization. The larger (10e,11o) active space converged successfully at the TS, but the calculations along the reaction pathway revealed that a Cr 3d orbital was replaced by a Cr 4s orbital at F03 and by a Cr 4d orbital at F09. The corresponding orbital similarities decreased to 0.915 and 0.923, respectively (Fig. 2c). These substitutions were identified by the \textbf{Consistency Auditor} and illustrate why a converged CASSCF calculation at a single geometry is not sufficient to establish the quality of an active space along a reaction pathway.
CASPER then tested a larger (14e,13o) active space that included two additional oxygen $\sigma$-donor orbitals. However, these orbitals were replaced by Cr 3p orbitals during the TS Run. The \textbf{Builder} subsequently proposed removing the fifth Cr 3d orbital, leading to a (10e,10o) active space that passed the orbital similarity test. Nevertheless, the \textbf{SA Watcher} identified changes in orbital character and abrupt variations in the NOONs between neighboring geometries. A three-root SA-CASSCF calculation indicated a crossing between electronic states in the F06--F08 region. CASPER therefore introduced state averaging over the two lowest triplet states.
The resulting (10e,10o) SA-CASSCF calculations passed the initial Path Review. However, the \textbf{Librarian} and \textbf{Reviewer} recommended including the fifth Cr 3d orbital to provide a more complete description of the metal d manifold. The final (10e,11o) SA-CASSCF active space satisfied the consistency criteria along the entire reaction pathway (Figure \ref{fig_rxnC}c). The corresponding NOs are shown in Supplementary Figure S5 and S6, and the energy profiles obtained with SA-CASSCF, NEVPT2, and L-PDFT are compared with $\omega$B97M-V in Figure \ref{fig_rxnC}d.

This example shows that CASPER can identify two different problems that may arise during multireference calculations. The first is the substitution of chemically relevant orbitals during CASSCF optimization, which may occur even when the calculation converges. The second is a change in the electronic-state character along the reaction pathway, which requires a different treatment of the electronic states. In Reaction C, both problems were identified and addressed without manual intervention.

\subsection{Reaction E: Maintaining a Chemically Meaningful Active Space}
\begin{figure}[t!]
  \includegraphics[width=0.9\textwidth]{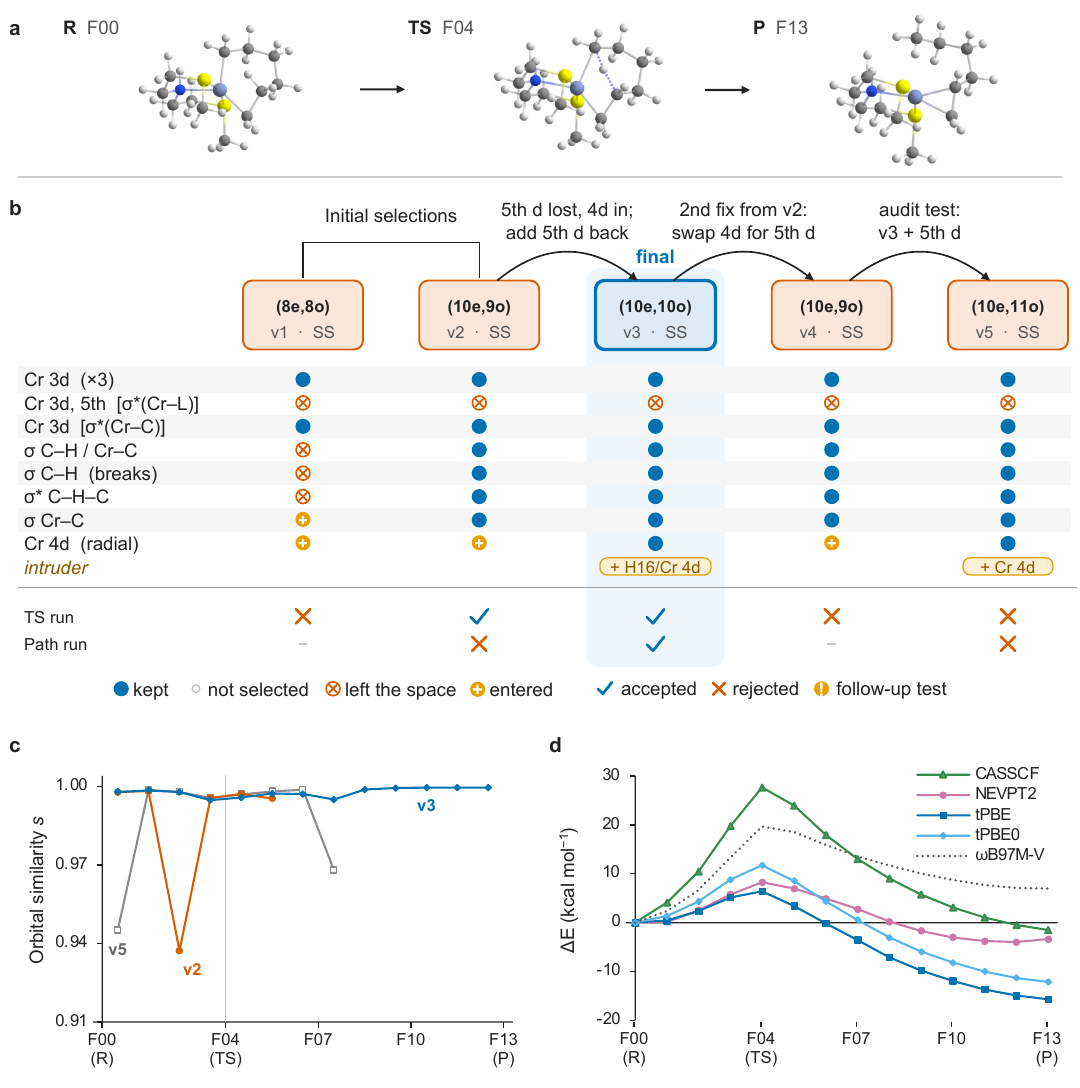}
  \caption{Step-by-step active space selection process along the path of Reaction E using CASPER. (a) Geometries for reactant, TS, and product, (b) The five trial active spaces in the order the workflow selects them. Each orbital row shows whether they are kept, or left or entered the active space during either TS or Path Run for each trial. The bottom two rows show the verdicts of the TS Run and the Path Scan. (c) Orbital similarity  between neighboring frames $s$ for the trials v2, v3 and v5. (d) Energy profile of the ground state of the adopted (10e, 10o) relative to reactant (F00) at state specific CASSCF, NEVPT2, MC-PDFT (tPBE, tPBE0), along with a respective $\omega$B97M-V result.}
  \label{fig_rxnE}
\end{figure}

Reaction E illustrates another difficulty in active space selection, where orbitals associated with the reacting \ce{C-H} bond and the Cr 3d manifold are readily replaced during CASSCF optimization (Figure \ref{fig_rxnE}). The initial (8e,8o) active space contained five Cr 3d orbitals and three orbitals associated with the bonds involved in hydrogen transfer. The \textbf{Chemist} proposed adding a \ce{Cr-C} $\sigma$  orbital, leading to a (10e,9o) active space.
During the TS calculation, a Cr 3d orbital associated with a \ce{Cr-ligand} $\sigma^*$ interaction was replaced by a Cr 4d orbital. The \textbf{Builder} therefore constructed a (10e,10o) active space to include the additional orbital while attempting to retain the original 3d orbital. However, during optimization, the same 3d orbital was again excluded and an orbital with mixed character involving the reacting hydrogen and Cr 4d entered the active space. Despite this substitution, the resulting active space remained consistent along the reaction pathway.
The \textbf{Chemist} accepted this selection because the excluded 3d orbital remained outside the active space at all geometries. However, the \textbf{Reviewer} questioned whether the missing orbital could affect the description of the metal electronic structure. CASPER therefore performed two additional tests. First, the (10e,9o) active space was reconstructed using the converged orbitals from the earlier calculation, but the same Cr 3d orbital was again replaced by a 4d orbital. Second, the excluded 3d orbital was added to the converged (10e,10o) selection, producing a (10e,11o) active space. This calculation also failed to retain the 3d orbital and did not satisfy the orbital consistency criterion (Figure \ref{fig_rxnE}c).

The \textbf{Chemist} therefore retained the (10e,10o) active space from the third trial. The repeated exclusion of the same Cr 3d orbital suggests that it does not contribute significantly to the static correlation described by this active space. The final NOs are shown in Figure S8. The corresponding CASSCF, NEVPT2, and MC-PDFT energy profiles are smooth along the reaction pathway (Figure \ref{fig_rxnE}d). However, compared with $\omega$B97M-V, NEVPT2 and MC-PDFT predict different relative reactant and product energies, favoring the product. These differences indicate that the choice of electronic structure method remains important even when a consistent active space has been identified.

Reaction E also shows that a larger active space is not necessarily a better one. The additional tests performed by CASPER did not improve the consistency of the calculations, and the smaller (10e,10o) space provided a more stable description along the reaction pathway. Importantly, the final selection was not based only on the convergence of the calculations, but also on the analysis of the orbital character and the results of the additional tests requested by the \textbf{Reviewer}.

\subsection{Comparison with Claude without the CASPER workflow}
To assess the contribution of the CASPER workflow, we repeated Reactions C and E using Claude Opus 5.5 without the specialized committee members and their instructions. Each control was performed in an independent Claude session with access to high-performance computing resources. The model received a short prompt specifying the reaction pathway, basis set, and objective of performing CASSCF calculations with a consistent active space and obtaining smooth potential energy profiles. The control environments were isolated from CASPER and from one another. The corresponding orbital selections, decision histories, orbital similarities, and energy profiles are provided in the Supplementary Information.

Claude alone was able to identify active spaces with high orbital consistency for both reactions, selecting (16e,14o) for Reaction C and (18e,14o) for Reaction E (Figures S1 and S2). However, both active spaces contained nearly inactive Cr 3s and 3p orbitals that were not needed to describe the relevant electronic structure. More importantly, for Reaction C, Claude did not identify the crossing between electronic states and therefore did not perform SA-CASSCF calculations. For Reaction E, the final active space did not contain the orbital associated with the reacting \ce{C-H} bond, which had been repeatedly replaced during CASSCF optimization. This omission may compromise the description of the electronic structure associated with hydrogen transfer.

These independent Claude sessions were able to obtain converged calculations and apparently consistent active spaces. However, the orbital analysis showed that important chemical features were not always properly described. In contrast, CASPER identified these problems and performed additional calculations to correct them. The comparison therefore shows the importance of providing the language model with a structured workflow, specific chemical criteria, and independent assessments of the results.

Although these control tests are limited to two reactions, they demonstrate that the ability to run electronic structure calculations is not sufficient for reliable autonomous multireference calculations. The additional chemical analysis and iterative assessment implemented in CASPER are essential for identifying problems that may otherwise remain unnoticed.

\section{Conclusion and Outlook}
We have developed CASPER, an agentic framework for automating multireference calculations along chemical reaction pathways. CASPER combines chemical knowledge, electronic structure calculations, and iterative assessment to select active spaces and ensure their consistency as the electronic structure changes along the reaction coordinate. We have tested CASPER on five transition-metal reactions, and in all cases it was able to identify consistent active spaces and generate smooth potential energy profiles with minimal human intervention.

An interesting finding of this work is that the independent session was able to perform the calculations and select active spaces that appeared consistent. However, a more careful analysis revealed that important aspects of the electronic structure were not properly described. In Reaction C, Claude did not recognize the crossing between electronic states and did not introduce state averaging. In Reaction E, it selected an active space that did not include orbitals involved in the reacting \ce{C-H} bond. CASPER was able to identify these problems and take the necessary steps to correct them. Although this comparison is limited to two reactions, it suggests that the structured workflow and the specialized instructions given to the agents play an important role in obtaining chemically meaningful results.

This is only the beginning. We plan to test CASPER on a much larger number of reactions, including more challenging systems involving multiple transition metals. In addition, we will investigate the performance of different language models to better understand the relative importance of the underlying model and the workflow. These studies will help us improve the efficiency and reliability of CASPER and extend its applicability to more complex chemical systems.

CASPER represents a promising first step towards making advanced multireference methods more accessible and useful for addressing challenging problems in chemistry and materials science. By analyzing the results, identifying possible problems, and suggesting how to improve the calculations, CASPER and other AI agents can help researchers gain a better understanding of the underlying physics and chemistry of the systems they are studying. We do not envision these agents replacing human expertise, but rather complementing it and allowing scientists to concentrate on the most important problems. There is still much to learn about how to make these workflows more reliable and efficient. 

\begin{acknowledgement}
This material is based upon work supported by the U.S. DOE, Office of Basic Energy Sciences, Division of Chemical Sciences, Geosciences, and Biosciences under Award Number USDOE/DE-SC0023382 and by the U.S. DOE, Office of Basic Energy Sciences, Heavy Elements Chemistry Program, under Award Number DE-SC0022572. 
M.R.H.~is partially supported by the National Science Foundation under Grant No. CHE-2513475. Developments of multiconfigurational methods by A.C.~were supported by the National Science Foundation under Grant No. CHE-2435218. L.G.~acknowledges all the above grants. 
The authors acknowledge the NVIDIA Academic Grant Program for providing GPU resources used in this work, the research computing center (RCC) at the University of Chicago
and National Energy Research Scientific Computing Center (NERSC). 
We acknowledge the use of Claude Opus 5.5 (Anthropic) in developing and implementing the agentic workflow presented in this work. Claude was also used to assist with the preparation and editing of the manuscript. All scientific results and conclusions were reviewed and validated by the authors, who take full responsibility for the content of this work.
We thank the Gagliardi lab members for their helpful discussion.
\end{acknowledgement}

\section*{Data Availability}
CASPER code, agent instructions, and run logs will be made publicly available upon acceptance.

\begin{suppinfo}
Results from independent Claude session without the CASPER workflow. Orbitals and occupation numbers of CASPER chosen active spaces for five test reactions.
\end{suppinfo}

\bibliography{refs}
\end{document}


\maketitle

\tableofcontents

\newpage

\section{Results from single prompt-based Claude test for reaction C and E}

\begin{figure}[H]
  \includegraphics[width=0.9\textwidth]{figures/SI_figures/case_C_control.pdf}
  \caption{Step-by-step active space selection process along the path of Reaction C with a control prompt with Claude. (a) Geometries for reactant, TS, and product, (b) The five trial active spaces in the order Claude selects them. Each orbital row shows whether they are kept, or left or entered the active space during either TS or Path Run for each trial. The bottom two rows show the verdicts of the TS Run and the Path Scan. (c) Orbital similarity between neighboring frames $s$ for the trials v1, v4 and v5. (d) Energy profile of the ground state of the adopted (16e,14o) SS relative to reactant (F00) at SA-CASSCF, MC-PDFT(tPBE, tPBE0), along with a respective $\omega$B97M-V result.}
  \label{fig_SI_rxnC}
\end{figure}

\begin{figure}[H]
  \includegraphics[width=0.9\textwidth]{figures/SI_figures/case_E_control.pdf}
  \caption{Step-by-step active space selection process along the path of Reaction E with a control prompt with Claude. (a) Geometries for reactant, TS, and product, (b) The four trial active spaces in the order Claude selects them. Each orbital row shows whether they are kept, or left or entered the active space during either TS or Path Run for each trial. The bottom two rows show the verdicts of the TS Run and the Path Scan. (c) Orbital similarity  between neighboring frames $s$ for the trials v1, v3 and v4. (d) Energy profile of the ground state of the adopted (18e,14o) SS relative to reactant (F00) at SA-CASSCF, MC-PDFT(tPBE, tPBE0), along with a respective $\omega$B97M-V result.}
  \label{fig_SI_rxnE}
\end{figure}

\begin{landscape}
\section{Orbitals and NOONs of each reactions from CASPER}

\begin{figure}[H]
  \centering
  \includegraphics[scale=0.68]{figures/SI_figures/reaction_A_orbitals.pdf}
  \caption{Selected (11e,11o) active space from CASPER for Reaction A.}
  \label{fig_CASPER_A}
\end{figure}
\end{landscape}

\begin{figure}[H]
  \includegraphics[scale=0.68]{figures/SI_figures/reaction_B_orbitals.pdf}
  \caption{Selected (10e,10o) active space from CASPER for Reaction B.}
  \label{fig_B}
\end{figure}

\begin{figure}[H]
\includegraphics[scale=0.68]{figures/SI_figures/reaction_C_orbitals.pdf}
  \caption{Selected (10e,11o) active space from CASPER for Reaction C. Shown is the state-specific NOs of the first root of SA(2)-CASSCF solution.}
  \label{fig_C0}
\end{figure}

\begin{figure}[H]
\includegraphics[scale=0.68]{figures/SI_figures/reaction_C_orbitals_root1.pdf}
  \caption{Selected (10e, 11o) active space from CASPER for Reaction C. Shown is the state-specific NOs of the second root of SA(2)-CASSCF solution.}
  \label{fig_C1}
\end{figure}

\begin{figure}[H]
    \includegraphics[scale=0.68]{figures/SI_figures/reaction_D_orbitals.pdf}
  \caption{Selected (14e, 16o) active space from CASPER for Reaction D.}
  \label{fig_D}
\end{figure}

\begin{figure}[H]
  \includegraphics[scale=0.68]{figures/SI_figures/reaction_E_orbitals.pdf}
  \caption{Selected (10e, 10o) active space from CASPER for Reaction E.}
  \label{fig_E}
\end{figure}

\begin{landscape}
\section{Orbitals and NOONs of each reaction from single prompt-based test}

\begin{figure}[H]
    \includegraphics[scale=0.68]{figures/SI_figures/reaction_C_orbitals_control.pdf}
  \caption{Selected (16e, 14o) active space from single prompt-based Claude test for Reaction C.}
  \label{fig_control_C}
\end{figure}

\begin{figure}[H]
    \includegraphics[scale=0.68]{figures/SI_figures/reaction_E_orbitals_control.pdf}
  \caption{Selected (18e, 14o) active space from single prompt-based Claude test for Reaction E.}
  \label{fig_control_E}
\end{figure}
\end{landscape}